\documentclass[conference,10pt]{IEEEtran}
\IEEEoverridecommandlockouts
\usepackage{cite}
\usepackage{amsmath,amssymb,amsfonts}
\usepackage{algorithmic}
\usepackage{algorithm}

\usepackage{graphicx}
\graphicspath{ {./} }
\usepackage{textcomp}
\usepackage{xcolor}
\usepackage{stfloats}
\usepackage{multirow}
\def\hrulefill{\leavevmode\leaders\hrule height 1pt\hfill\kern0pt}
\makeatletter
\def\BibTeX{{\rm B\kern-.05em{\sc i\kern-.025em b}\kern-.08em
    T\kern-.1667em\lower.7ex\hbox{E}\kern-.125emX}}
\newcounter{MYtempeqncnt}

\makeatletter

\begin{document}

\title{Distributed Resource Block Allocation for Wideband Cell-free System
\vspace{-0.0cm}
}

\author{\IEEEauthorblockN{Yang Ma, Shengqian Han, and Chenyang Yang}
	\IEEEauthorblockA{School of Electronic and Information Engineering, Beihang University, Beijing 100191, China\\
		Email: \{yangma, sqhan, cyyang\}@buaa.edu.cn}
\vspace{-0.7cm}
}

\maketitle

\begin{abstract}
This paper studies distributed resource block (RB) allocation in wideband orthogonal frequency-division multiplexing (OFDM) cell-free systems. We propose a novel distributed sequential algorithm and its two variants,  which optimize RB allocation based on the information obtained through over-the-air (OTA) transmissions between access points (APs) and user equipments, enabling local decision updates at each AP. To reduce the overhead of OTA transmission, we further develop a distributed deep learning (DL)-based method to learn the RB allocation policy. Simulation results demonstrate that the proposed distributed algorithms perform close to the centralized algorithm, while the DL-based method outperforms existing baseline methods.

\end{abstract}


\section{Introduction}

Coordinated resource block (RB) allocation among access points (APs) is crucial for cell-free networks. On one hand, allocating RBs to the user equipments (UEs) serves as a prerequisite for other resource allocation schemes, such as the widely studied coordinated beamforming~\cite{elina2017precoding, Dujoint2021}. On the other hand, uncoordinated RB allocation among APs can convert signals into interference, leading to significant performance~degradation.

Existing studies have investigated the centralized RB allocation for cell-free networks. In \cite{Zhang2021precoding}, a numerical algorithm was proposed for a wideband reconfigurable intelligent surface-aided cell-free system to optimize subcarrier allocation. In \cite{adam2024learning}, a learning-based method was proposed to optimize RB allocation for cell-free networks. Both studies assumed that APs are connected to a central processing unit (CPU) via fronthaul links, which optimizes the resource allocation policy and subsequently conveys the allocation decisions to APs. However, centralized optimization incurs huge fronthaul overhead and computational complexity as the system size scales up~\cite{elina2017precoding, Dujoint2021, Zhang2021precoding, adam2024learning}. Moreover, in real-world systems, APs may be controlled by different CPUs, making real-time information exchange via fronthaul links~impractical.

The limitations of centralized optimization call for distributed optimization, which is anticipated to play a key role in future wireless systems \cite{ITU2030}. Distributed optimization exhibits a trade-off between performance and scalability \cite{Emil2020Making, Italo2021distributed, Ammar2022Distributed, Lee2024fully}. Fully distributed beamforming algorithms for single-carrier systems have been proposed in \cite{Italo2021distributed} and \cite{Ammar2022Distributed}.
To the best of our knowledge, only \cite{Lee2024fully} addressed subcarrier allocation for a distributed wideband orthogonal frequency-division multiplexing (OFDM) cell-free system. However, this work still assumes a fronthaul network for information sharing among the primary AP and secondary APs within each cluster.

Most existing works on coordinated distributed resource allocation for cell-free networks employ iterative numerical algorithms \cite{Italo2021distributed, Ammar2022Distributed}, which can achieve satisfactory performance through iterative computation but often struggle to meet real-time requirements. Deep learning (DL)-based methods can tackle complex optimization problems while offering low inference complexity. In \cite{wang2023rui} and \cite{chen2024distributed}, DL-based methods were proposed for distributed optimization in cell-free systems, both focusing on optimizing the beamforming policy for single-carrier systems. Nevertheless, a distributed numerical or DL-based method for the optimization of RB allocation in wideband cell-free systems is still lacking.

In this paper, we study distributed RB allocation in a wideband cell-free system, aiming to maximize the sum rate of the system. A distributed sequential algorithm and its two variants are proposed to optimize the RB allocation, requiring no fronthaul links for information exchange among APs. Instead, they exploit the information obtained through over-the-air (OTA) transmission between APs and UEs to update decisions locally at each AP. To reduce the signaling overhead of OTA transmission, we further propose a distributed DL-based method to learn the RB allocation policy.
Simulation results demonstrate that the proposed distributed algorithms perform close to the centralized algorithm, and the DL-based method outperforms existing baseline methods.

\vspace{-0.1cm}
\section{System Model}
Consider a time division duplexing OFDM cell-free system, where $N$ single-antenna APs serve $K$ single-antenna UEs on $F$ orthogonal RBs.
Each RB consists of $C$ consecutive subcarriers, experiencing the same channels.

The signal transmitted by AP$_n$ can be expressed as ${x}_{n,f_c}=\sum_{k=1}^K{v_{n,k,f}}{S}_{k,f_c}$, where $S_{k,f_c}\sim \mathcal{CN}(0, 1)$ is the data to UE$_k$ on subcarrier $f_c$ within RB$_f$, $c=1,2,\cdots, C$, and $v_{n,k,f} \in \mathbb{C}$ is the decision variable of AP$_n$ used to weight the data for UE$_k$. The term $|v_{n,k,f}|^2$ denotes the power allocated by AP$_n$ to UE$_k$ on RB$_f$. If $|v_{n,k,f}|^2$ is zero, then RB$_f$ is not allocated to UE$_k$ at AP$_n$.

The received signal at UE$_k$ on RB$_f$ is
\begin{equation}\label{received_signal}
	\begin{split}
		{y}_{k,f} &=\sum_{n=1}^Nh_{n,k,f}{x}_{n,f_c}+{\rho}_{k, f}=\sum_{n=1}^N{h_{n,k,f}}v_{n,k,f}{S}_{k,f_c}   \\
		&+\sum_{j=1,j\ne k}^K\sum_{n=1}^N{h_{n,k,f}}{v_{n,j,f}}{S}_{j,f_c}+{\rho}_{k, f},
	\end{split}
\end{equation}
where $h_{n,k,f} = \sqrt{\beta_{n, k}}\tilde  {h}_{n,k,f}\in \mathbb{C}$ is the channel from AP$_n$ to UE$_k$ on RB$_f$, $\beta_{n,k}$ and $\tilde  {h}_{n,k,f}$ denote the large-scale and small-scale channels, respectively, ${\rho}_{k, f}\sim\mathcal{CN}(0, \sigma^2_{k,f})$ is the noise at UE$_k$ on RB$_f$, and $\sigma^2_{k,f}$ is the noise power on RB$_f$. 

The signal-to-interference-plus-noise (SINR) of UE$_k$ on RB$_f$ is
\begin{equation}\label{SINR}
	\mathrm{SINR}_{k,f}=\frac{\lvert G_{k,f}^{k}\rvert ^2}{\sum_{j\neq k}\lvert G_{k,f}^{j}\rvert ^2+\sigma^2_{k,f}},
\end{equation}
where $G_{k,f}^{k'}=\sum_{n=1}^N{h_{n,k,f}}v_{n,k',f}, k'=1,\cdots, K$.
The sum rate of UEs can be obtained as
\begin{equation}\label{UE_rate_perRB}
	\mathrm{SR}=C\sum_{f=1}^{F}\sum_{k=1}^{K}\log _2\left( 1+\mathrm{SINR}_{k,f}\right).
\end{equation}

The RB allocation problem can be formulated as
\begin{subequations}\label{Optimization_problem}
\begin{align}
	\text{P1:}\quad \max_{\mathbf{V}}\ &  \quad\mathrm{SR} \\
	s.t.\quad & \lVert\mathbf{V}_n\rVert_\mathsf{F}^2\leq P_\mathrm{t},\ n=1,2,\cdots, N,\label{power_constraint}
\end{align}
\end{subequations}
where $\mathbf{V}_n\in\mathbb{C}^{K\times F}$ with $v_{n,k,f}$ being the $(k,f)$-th element, $\mathbf{V}=[\mathbf{V}_1, \cdots,\mathbf{V}_N]$, and $P_\mathrm{t}$ is the maximum transmit power of each AP.

\section{Distributed Algorithms}

In this section, we first briefly introduce the centralized weighted sum mean squared error minimization (WMMSE) algorithm for solving P1. Then, we propose a distributed sequential WMMSE (SWMMSE) algorithm and its two variants.

\subsection{Centralized WMMSE Algorithm}
According to the analysis in~\cite{shi2011iteratively}, problem P1 is equivalent to the following problem
\begin{IEEEeqnarray*}{rl}\IEEEyesnumber\label{Optimization_problemP2}
	\text{P2:} \quad \min_{\mathbf{V}}\ &  \sum_{k=1}^{K}\sum_{f=1}^{F}w_{k, f}\varepsilon_{k,f}(\mathbf{V})\\
	s.t. \ & (\ref{power_constraint}),
\end{IEEEeqnarray*}
where $\varepsilon_{k,f}$ is the mean squared error (MSE) of UE$_k$ on RB$_f$, which can be expressed as
\begin{equation}\label{MSE}
	\begin{split}
		&\varepsilon_{k,f}(\mathbf{V}) =\mathbb{E}(\lvert U_{k,f}y_{k,f}-S_{k,f_c}\rvert^2)   \\
		&=|U_{k,f}G_{k,f}^k-1|^2+\sum_{j\neq k}|U_{k,f}G_{k,f}^j|^2+|U_{k,f}|^2\sigma_{k,f}^2,
	\end{split}
\end{equation}
where $U_{k,f}$ and $w_{k, f}$ are auxiliary variables interpreted as the receiving coefficient and the weight coefficient of UE$_k$ on RB$_f$, respectively.

The WMMSE algorithm alternatingly updates $v_{n,k,f}$, $U_{k,f}$, and $w_{k, f}$ given any two of them. Specifically, $U_{k,f}$ and $w_{k, f}$ are updated as
\begin{gather}
	U_{k,f}=\frac{{G_{k,f}^{k*}}}{{\left| G_{k,f}^k \right|^2+\sum_{j\neq k}|G_{k,f}^j|^2+\sigma^2_{k,f}}},\label{combining1}\\
    w_{k, f}=\frac{1}{\varepsilon_{k,f}}.\label{combining2}
\end{gather}
Given $U_{k,f}$ and $w_{k, f}$, problem P2 is convex for $v_{n,k,f}$, and its optimal solution can be readily obtained.

When used in a cell-free system, the WMMSE algorithm requires a CPU to frequently collect channel information from APs and convey decisions to APs through fronthaul links, resulting in significant fronthaul overhead. In the next subsection, we propose the SWMMSE algorithm, which enables each AP to locally update its decisions without requiring fronthaul links for information exchange.

\subsection{Distributed Sequential WMMSE Algorithm}\label{SeqWMMSE_algo}

\begin{figure}[ht]
\centering
	\includegraphics[width=0.8\linewidth]{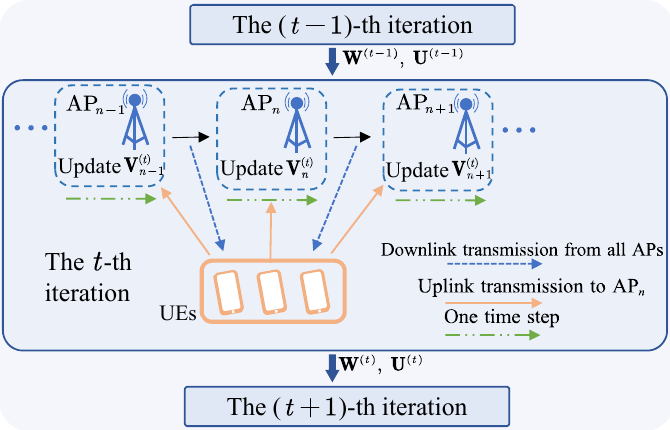}
	\caption{Procedure of the SWMMSE algorithm.}
	\label{Seq_WMMSE}
\end{figure}

\subsubsection{\underline{Local Decision of RB Allocation}}\label{procedure_SWMMSE}
 The SWMMSE algorithm lets APs update their decisions in a sequential fashion. In each time step, only one AP updates its decision while the other APs retain their previous decisions. The procedure of the SWMMSE algorithm is illustrated in~Fig.~\ref{Seq_WMMSE}.

Each iteration of the algorithm is comprised of $N$ time steps, with one AP updating its decision in each time step. Specifically, in the $n$-th time step of the $t$-th iteration, AP$_1$ to AP$_{n-1}$ have already updated their decisions, denoted by $\mathbf{V}_{n<}^{(t)}\triangleq[\mathbf{V}_1^{(t)}, \cdots, \mathbf{V}_{n-1}^{(t)}]$, while AP$_{n+1}$ to AP$_{N}$ retain their decisions from the $(t\!-\!1)$-th iteration, denoted by $\mathbf{V}_{n>}^{(t-1)}\triangleq[\mathbf{V}_{n+1}^{(t\!-\!1)}, \cdots, \mathbf{V}_N^{(t\!-\!1)}]$. In this time step, only AP$_n$ updates its decision by solving the following local optimization problem
\begin{subequations}\label{Optimization_problemP4}
\begin{align}
	\text{P3:} \quad \min_{\mathbf{V}_n^{(t)}}\ & \sum_{k, f}w_{k,f}^{(t\!-\!1)}\varepsilon_{k,f}\Big(\mathbf{V}_{n<}^{(t)}, \mathbf{V}^{(t)}_{n},\mathbf{V}_{n>}^{(t-1)}\Big)\\
	\quad s.t.\quad & \lVert\mathbf{V}_n\rVert_\mathsf{F}^2\leq P_\mathrm{t},\label{Optimization_problemP4cc}
\end{align}
\end{subequations}
where only $\mathbf{V}_n^{(t)}$ is optimized and $w_{k,f}^{(t\!-\!1)}$ is the coefficient obtained in the $(t\!-\!1)$-th~iteration.

Similar to \eqref{MSE}, the MSE is expressed as
\begin{equation}\label{SWMMSE_MSE}
	\begin{aligned}
	\varepsilon_{k,f}\left(\cdot\right)=&|U_{k,f}^{(t\!-\!1)}G_{n,k,f}^{k(t)}(\mathbf{V}_{n\leq}^{(t)},\mathbf{V}_{n>}^{(t-1)})-1|^2+|U_{k,f}^{(t\!-\!1)}|^2\sigma^2_{k,f}\\
& \ \ +\sum_{j\neq k}|U_{k,f}^{(t\!-\!1)}G_{n,k,f}^{j(t)}(\mathbf{V}_{n\leq}^{(t)},\mathbf{V}_{n>}^{(t-1)})|^2,
	\end{aligned}
\end{equation}
where $U_{k,f}^{(t\!-\!1)}$ is the receiving coefficient obtained in the $(t\!-\!1)$-th iteration. According to the definition of $G_{k,f}^{k'}$ provided below \eqref{SINR}, $G_{n,k,f}^{k'(t)}(\mathbf{V}_{n\leq}^{(t)},\mathbf{V}_{n>}^{(t-1)})$ can be expressed as
\begin{equation} \label{E:G1}
	\begin{split}
G_{n,k,f}^{k'(t)}(\cdot)=&\underbrace{\sum_{m=1}^{n-1}h_{m,k,f}v_{m,k',f}^{(t)}+\sum_{m'=n}^{N}h_{m',k,f}v_{m',k',f}^{(t\!-\!1)}}_{\triangleq\widetilde{G}_{n,k,f}^{k'(t)}}\\
                &\ \ - h_{n,k,f}v_{n,k',f}^{(t\!-\!1)} + h_{n,k,f}v_{n,k',f}^{(t)}\\
                =& \widetilde{G}_{n,k,f}^{k'(t)}- h_{n,k,f}v_{n,k',f}^{(t\!-\!1)} + h_{n,k',f}v_{n,k',f}^{(t)}.
	\end{split}
\end{equation}
Problem P3 can be solved by examining its Karush-Kuhn-Tucker conditions. We can derive a closed-form expression for the decision $v_{n,k,f}^{(t)}$, as given by (\ref{how2feedback}) at the bottom of this page, where $\mu_n$ is the Lagrangian multiplier that ensures constraint \eqref{Optimization_problemP4cc}.

\begin{figure*}[!b]
	\normalsize
	\setcounter{MYtempeqncnt}{\value{equation}}
	\hrulefill
	\setcounter{equation}{11}

	\begin{equation}\label{how2feedback}
		\begin{aligned}
			v_{n,k,f}^{(t)}= & \frac{1}{\mu _n+\underbrace{\sum_{i=1}^{K}{w_{i,f}^{(t\!-\!1)}|h_{n,i,f}|^2|U_{i,f}^{(t\!-\!1)}|^2}}_{\triangleq D_{n,f}^{(t)}}}
\bigg(\underbrace{h_{n,k,f} w_{k,f}^{(t\!-\!1)}U_{k,f}^{(t\!-\!1)}}_{\triangleq A_{n,k,f}^{(t)}} + \underbrace{\sum_{i=1}^{K}w_{i,f}^{(t\!-\!1)}|h_{n,i,f}|^2|U_{i,f}^{(t\!-\!1)}|^2}_{\triangleq D_{n,f}^{(t)}}v_{n,k,f}^{(t\!-\!1)*} - \\
  &\qquad \underbrace{h_{n,k,f}\cdot w_{k,f}^{(t\!-\!1)}|U_{k,f}^{(t\!-\!1)}|^2 \widetilde{G}_{n,k,f}^{k*}+\sum_{j\ne k}{ h_{n,j,f}\cdot w_{j,f}^{(t\!-\!1)}|U_{j,f}^{(t\!-\!1)}|^2 \widetilde{G}_{n,j,f}^{k*}}}_{\triangleq M_{n,k,f}^{(t)}}\bigg)^\ast
		\end{aligned}
	\end{equation}
	\setcounter{equation}{\value{MYtempeqncnt}}
\end{figure*}
\setcounter{equation}{12}

In (\ref{how2feedback}), three terms, namely $A_{n,k,f}^{(t)}$, $D_{n,f}^{(t)}$, and $M_{n,k,f}^{(t)}$, are defined. As part of the distributed SWMMSE algorithm, AP$_n$ should locally acquire these terms to update its RB allocation decision with (\ref{how2feedback}). The challenge for an AP in making decisions in a distributed manner arises from its lack of awareness of the decisions made by other APs. The proposed algorithm addresses this issue by enabling information sharing via the OTA transmission between APs and UEs. 

\subsubsection{\underline{Information Sharing via OTA Transmission}}\label{information_exchange}

OTA transmission occurs in each time step, which can be divided into two phases: downlink OTA transmission and uplink OTA transmission. After these two phases, each AP can acquire $A_{n,k,f}^{(t)}$, $D_{n,f}^{(t)}$, and $M_{n,k,f}^{(t)}$ to update its own decision in the same time step.

$\bullet$ \emph{Downlink OTA transmission: }
The $n$-th time step in the $t$-th iteration begins with the transmission of downlink pilots from all APs. The pilots are precoded with the current RB allocation decisions $\mathbf{V}_{n<}^{(t)}$ and $\mathbf{V}_{n\geq}^{(t-1)}$.
The transmitted pilot signal of AP$_m$ on RB$_f$ is 
$\eta_{\mathsf{d}}\sum_{k=1}^K v_{m,k,f}^{(t)} \mathbf{s}_{k}^{\mathsf{p}}$ for $m<n$ and $\eta_{\mathsf{d}}\sum_{k=1}^K v_{m,k,f}^{(t\!-\!1)} \mathbf{s}_{k}^{\mathsf{p}}$ for $m\ge n$, where $\eta_{\mathsf{d}}$ is a scaling factor ensuring the maximum power constraint of every AP, $\mathbf{s}_{k}^{\mathsf{d}}$ is the downlink pilot sequence of UE$_k$ satisfying $\|\mathbf{s}_{k}^{\mathsf{p}}\|^2_{\mathsf{F}}=1$, and $\mathbf{s}_{k}^{\mathsf{p}}$ and $\mathbf{s}_{j}^{\mathsf{p}}$ are orthogonal for $j\ne k$.
The received pilot signal at UE$_k$ on RB$_f$ can be expressed as
\begin{equation}\label{downlink_pilot_signal}	\mathbf{y}^{\mathsf{p}}_{k,f}=\eta_{\mathsf{d}}\widetilde{G}_{n,k,f}^{k(t)}\mathbf{s}_{k}^{\mathsf{p}}
+\eta_{\mathsf{d}}\sum_{j\ne k}\widetilde{G}_{n,k,f}^{j(t)}\mathbf{s}_{j}^{\mathsf{p}}, 
\end{equation}
where $\widetilde{G}_{n,k,f}^{k'(t)}$ is defined in \eqref{E:G1}, and the noise is omitted for clarity. 

From \eqref{downlink_pilot_signal}, UE$_k$ can estimate $\widetilde{G}_{n,k,f}^{(t)}$ as $\frac{1}{\eta_{\mathsf{d}}}(\mathbf{s}_{k}^{\mathsf{p}})^H\mathbf{y}^{\mathsf{p}}_{k,f}$ and $\sum_{j\ne k}|\widetilde{G}_{n,k,f}^{j(t)}|^2$ as $\frac{1}{\eta_{\mathsf{d}}^2} \|\mathbf{y}^{\mathsf{p}}_{k,f}-\eta_{\mathsf{d}}
\widetilde{G}_{n,k,f}^{(t)}\mathbf{s}_{k}^{\mathsf{p}}\|^2_{\mathsf{F}}$ in every time step. Once $N$ time steps in an iteration are completed, UE$_k$ updates the receiving coefficient $U_{k,f}^{(t)}$ and the weight coefficient $w_{k,f}^{(t)}$ using \eqref{combining1}, \eqref{combining2} and \eqref{MSE}, where
${G_{k,f}^{k'}}$ is replaced with $\widetilde{G}_{n,k,f}^{k'(t)}$.

\begin{figure}[htbp]
	\centerline{\includegraphics[width=0.8\linewidth]{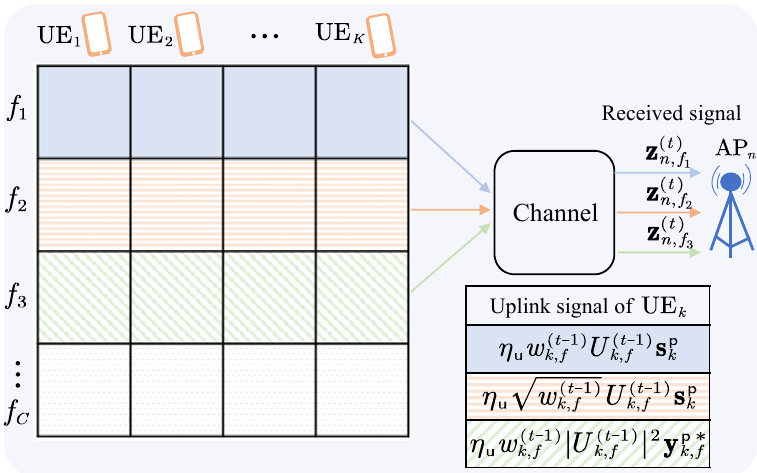}}\vspace{0mm}
	\caption{Illustration of uplink OTA transmission.}
	\label{pilot_pattern_perRB}
\end{figure}

$\bullet$ \emph{Uplink OTA transmission: } The UEs transmit different pilot signals, as shown in~Fig.~\ref{pilot_pattern_perRB}, to facilitate each AP in locally estimating $A_{n,k,f}^{(t)}$, $D_{n,f}^{(t)}$, and $M_{n,k,f}^{(t)}$.

\begin{itemize}
  \item[$\mathrm{i.}$] Acquisition of $A_{n,k,f}^{(t)}$: With the coefficients $w_{k,f}^{(t\!-\!1)}$ and $U_{k,f}^{(t\!-\!1)}$ obtained in the $(t\!-\!1)$-th iteration, each UE$_k$ transmits the pilot signal $\eta_{\mathsf{u}} w_{k,f}^{(t\!-\!1)}U_{k,f}^{(t\!-\!1)}\mathbf{s}_{k}^{\mathsf{p}}$ on subcarrier $f_1$, where $\eta_{\mathsf{u}}$ is a scaling factor ensuring the maximum power constraint of every UE. Denoting $\mathbf{z}_{n,f_1}^{(t)}$ as the received signal at AP$_n$, $A_{n,k,f}^{(t)}$ can be estimated as $\frac{1}{\eta_{\mathsf{u}}} (\mathbf{s}_{k}^{\mathsf{p}})^H\mathbf{z}^{(t)}_{n,f_1}$.
  \item[$\mathrm{ii.}$] Acquisition of $D_{n,f}^{(t)}$: Each UE$_k$ transmits the pilot signal $\eta_{\mathsf{u}} \sqrt{w_{k,f}^{(t\!-\!1)}}U_{k,f}^{(t\!-\!1)}\mathbf{s}_{k}^{\mathsf{p}}$ on subcarrier $f_2$. Denoting $\mathbf{z}_{n,f_2}^{(t)}$ as the received signal at AP$_n$,  $D_{n,f}^{(t)}$ can be estimated as $\frac{1}{\eta_{\mathsf{u}}^2} \|\mathbf{z}^{(t)}_{n,f_2}\|^2_{\mathsf{F}}$.
  \item[$\mathrm{iii.}$] Acquisition of $M_{n,k,f}^{(t)}$: Each UE$_k$ transmits the signal $\eta_{\mathsf{u}} w_{k,f}^{(t\!-\!1)}|U_{k,f}^{(t\!-\!1)}|^2 \mathbf{y}^{\mathsf{p}*}_{k,f}$ on subcarrier $f_3$, where $\mathbf{y}^{\mathsf{p}*}_{k,f}$ is the conjugated received downlink signal as given in \eqref{downlink_pilot_signal}. Denoting $\mathbf{z}_{n,f_3}^{(t)}$ as the received signal at AP$_n$, $M_{n,k,f}^{(t)}$ can be estimated as $\frac{1}{\eta_{\mathsf{u}}} (\mathbf{s}_{k}^{\mathsf{p}})^H\mathbf{z}^{(t)}_{n,f_3}$.
\end{itemize}

With the acquired $A_{n,k,f}^{(t)}$, $D_{n,f}^{(t)}$, and $M_{n,k,f}^{(t)}$, AP$_n$ can locally compute $v_{n,k,f}^{(t)}$ using (\ref{how2feedback}) and update the decision as
\begin{equation}\label{step_update}
	{v}_{n,k,f}^{(t)}\leftarrow\gamma^{(t)}{v}_{n,k,f}^{(\! t-1\!)} + (1-\gamma^{(t)}){v}_{n,k,f}^{(t)},
\end{equation}
\noindent where $\gamma^{(t)}\in[0,1)$ is the step size used for the $t$-th iteration to guarantee convergence. The procedure of SWMMSE is summarized in Algorithm \ref{SMMSE}.

\begin{algorithm}[htbp]
	\caption{The SWMMSE Algorithm}
	\label{SMMSE}
	\begin{algorithmic}[1]
		\STATE {Randomly initialize $\mathbf{V}^{(0)}=[\mathbf{V}^{(0)}_1,\cdots,\mathbf{V}^{(0)}_N]$. Set the tolerance of accuracy $\epsilon$, and $t=0$.}
		\WHILE{$\lVert\mathbf{V}^{(t)}-\mathbf{V}^{(t\!-\!1)}\rVert_\mathsf{F} > \epsilon$}
		\STATE {$t\leftarrow t+1$.}
        \STATE {UEs compute $\mathbf{W}^{(t\!-\!1)}$ and $\mathbf{U}^{(t\!-\!1)}$.}
		\FOR{$n=1, 2, \cdots, N$}
		\STATE {All APs transmit precoded pilot signals in the downlink OTA transmission, as specified in Sec.~\ref{information_exchange}.}
		\STATE {All UEs transmit precoded pilot signals in the uplink OTA transmission, as specified in Sec.~\ref{information_exchange}.}
		\STATE {Each AP$_n$ estimates $A_{n,k,f}^{(t)}, M_{n,k,f}^{(t)}$, and $D_{n,f}^{(t)}$ and updates $\mathbf{V}_n^{(t)}$ locally using (\ref{how2feedback}) and \eqref{step_update}.}
        \ENDFOR        
        \ENDWHILE
	\end{algorithmic}
        
\end{algorithm}

\subsection{Two Variants of SWMMSE Algorithm}

The SWMMSE algorithm updates the decision of APs one by one, leading to high OTA transmission overhead. To reduce the overhead, we introduce two variants of SWMMSE. 

The first variant, referred to as PWMMSE, updates RB decisions of all APs simultaneously. Compared to SWMMSE, PWMMSE achieves faster convergence and lower OTA transmission overhead. A similar update strategy is employed in \cite{Italo2021distributed}. For PWMMSE, the RB decision can still be obtained using \eqref{how2feedback} and \eqref{step_update}, except that the term $\widetilde{G}_{n,k,f}^{k'(t)}$ defined in \eqref{E:G1} is replaced by $\widetilde{G}_{n,k,f}^{k'(t)} = \sum_{n=1}^{N} h_{n,k,f}{v}_{n,k',f}^{(t\!-\!1)}$.

The second variant provides a trade-off between PWMMSE and SWMMSE, in which $N$ APs are clustered into $Q$ clusters, and the APs within the same cluster to update their decisions simultaneously at each time step. The resulting algorithm is referred to as C-SWMMSE. Each iteration of C-SWMMSE is comprised of $Q$ time steps. In the $q$-th time step of the $t$-th iteration, the APs in cluster $q$ update their decisions. Denote $\mathcal{C}_{q}$ as the index set of the APs in cluster $q$. For AP$_n$  in  $\mathcal{C}_{q}$, $\widetilde{G}_{n,k,f}^{k'(t)}$ is given by
\begin{equation}\label{E:G2}
	\begin{split}
		\widetilde{G}_{n,k,f}^{k'(t)}=&\sum_{i=1}^{q-1}\sum_{m\in\mathcal{C}_{i}} h_{m,k,f}v_{m,k',f}^{(t)}+
		\sum_{n\in\mathcal{C}_{q}}h_{n,k,f}v_{n,k',f}^{(t\!-\!1)}\\
		&\ +\sum_{j=q+1}^{Q} \sum_{m'\in\mathcal{C}_{j}}h_{m',k,f}v_{m',k',f}^{(t\!-\!1)}.
	\end{split}
\end{equation}
The term $\widetilde{G}_{n,k,f}^{k'(t)}$ of PWMMSE (C-SWMMSE) indicates that AP$_n$ is only aware of the outdated decisions $v_{n,k,f}^{(t\!-\!1)}$ of other APs (within the same cluster) from the previous iteration. Such outdated information may lead to performance degradation. The convergence of SWMMSE, C-SWMMSE, and PWMMSE can be established using the best response analysis presented in \cite{Scutari2014dec}, thus the detailed convergence analysis is omitted due to the space limitation.

\section{Distributed Deep Learning Method}

For the purpose of harnessing the benefits of PWMMSE in reducing signaling overhead, where all APs simultaneously update their decisions, we propose a distributed DL-based method (DDM). 

The core idea of DDM  is to let deep neural networks (DNNs) learn to infer local RB decisions $\mathbf{V}_n$ using the information obtained via OTA transmission based on PWMMSE. However, unlike traditional deep unfolding methods \cite{hu2021deepunfold, Schynol2023coordinated}, where the output of a sub-DNN is directly fed into the next, the proposed DDM incorporates an OTA block between the sub-DNNs of two adjacent time steps to facilitate distributed optimization. Furthermore, instead of employing different sub-DNNs to learn key variables of a numerical algorithm, the information obtained via OTA transmission is directly fed into the sub-DNNs. This design enables the sub-DNNs to learn efficient allocation decisions when the number of time steps is very limited.

\begin{figure}[htbp]
	\centerline{\includegraphics[width=0.9\linewidth]{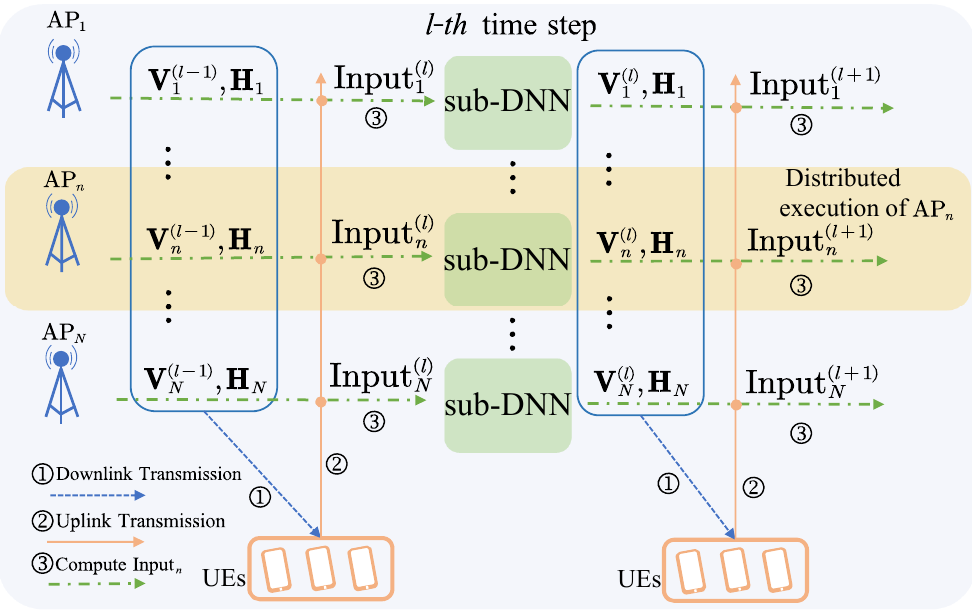}}\vspace{0mm}
	\caption{The illustration of DDM in the $l$-th time step. The shaded area represents the distributed execution procedure of each AP$_n$.}
	\label{Para_DNN_Framework}
\end{figure}

To obtain the decisions of all APs, the DDM requires a total of $N\times L$ sub-deep neural networks (DNNs), corresponding to $N$ APs and $L$ time steps. The $N$ sub-DNNs for the same time step share parameters (i.e., weight matrices and biases of the DNN), while the sub-DNNs for different time steps are different. The procedure of DDM in the $l$-th time step is shown in Fig.~\ref{Para_DNN_Framework}. Specifically, in the $l$-th time step, $N$ APs first conduct downlink OTA transmission using the output of sub-DNNs from the $(l\!-\!1)$-th step, denoted by $\mathbf{V}_1^{(l\!-\!1)}, \cdots, \mathbf{V}_N^{(l\!-\!1)}$. Subsequently, $K$ UEs compute their receiving and weight coefficients $w_{k,f}^{(l\!-\!1)}\in\mathbf{W}^{(l\!-\!1)}$ and $U_{k,f}^{(l\!-\!1)}\in\mathbf{U}^{(l\!-\!1)}$, and then conduct uplink OTA transmission as described in Sec.~\ref{information_exchange}. Consequently, each AP$_n$ can acquire $A_{n,k,f}^{(l)}$, $D_{n,f}^{(l)}$, and $M_{n,k,f}^{(l)}$ as done in Sec.~\ref{information_exchange} as
\begin{gather}
  A_{n,k,f}^{(l)}=h_{n,k,f}w_{k,f}^{(l\!-\!1)}U_{k,f}^{(l\!-\!1)}, \\ D_{n,f}^{(l)}=\sum_{k=1}^{K}{w_{k,f}^{(l\!-\!1)}|h_{n,k,f}|^2|U_{k,f}^{(l\!-\!1)}|^2},\\
  	\begin{aligned}
M_{n,k,f}^{(l)}=&
  h_{n,k,f}w_{k,f}^{(l\!-\!1)}|U_{k,f}^{(l\!-\!1)}|^2 \widetilde{G}_{n,k,f}^{k(l\!-\!1)*}\\
  & \ +\sum_{j\ne k}{ h_{n,j,f}w_{j,f}^{(l\!-\!1)}|U_{j,f}^{(l\!-\!1)}|^2 \widetilde{G}_{n,j,f}^{k(l\!-\!1)*}},
	\end{aligned}
\end{gather}
where $\widetilde{G}_{n,k,f}^{k'(l\!-\!1)}=\sum_{n=1}^{N}h_{n,k,f}v_{n,k',f}^{(l\!-\!1)}$, $k'=1,\cdots,K$.

The information acquired via the OTA transmissions forms the input of the sub-DNNs for the $l$-th time step. For the $n$-th sub-DNN, its input is represented as $\mathrm{Input}_{n}^{(l)}=\{\Re(\mathbf{R}_{n}^{(l)}),\Im(\mathbf{R}_{n}^{(l)}),\mathbf{B}_{n}^{(l)}\}$, where $\mathbf{R}_{n}^{(l)}$ is a $K\times F$ matrix, with the $(k,f)$-th element defined as
\begin{equation}\label{DDM_input}	R_{n,k,f}^{(l)}\triangleq\left(A_{n,k,f}^{(l)}-M_{n,k,f}^{(l)}\right)^*.
\end{equation}
In addition, we can observe from (\ref{how2feedback}) that the first term on the right-hand side involves $D_{n,f}^{(l)}$. Thus, we set $\mathbf{B}_{n}^{(l)}\triangleq\mathbf{1}_{K}\mathbf{d}_n\in\mathbb{R}^{K\times F}$, where $\mathbf{d}_n=[\hat{D}_{n,1}^{(l)},\dots, \hat{D}_{n,F}^{(l)}]$ and $\mathbf{1}_{K}$ denotes the all-one vector of dimension $K$. 

The output of the $n$-th sub-DNN for the $l$-th time step is given by $\{\Re(\mathbf{V}_n^{(l)}), \Im(\mathbf{V}_n^{(l)})\}\in\mathbb{R}^{K\times F\times 2}$. From the output, we can obtain the decision $\mathbf{V}_n^{(l)}$, which is then scaled by $\sqrt{P_{\mathrm{t}}}/\|\mathbf{V}_n^{(l)}\|_{\mathsf{F}}$ to satisfy the per-AP power constraint. Before the first time step, the initial decisions (i.e., $\mathbf{V}_1^{(0)}, \cdots, \mathbf{V}_N^{(0)}$) can be randomly initialized. With the scaled decision $\mathbf{V}_n^{(l)}$, AP$_n$ updates its decision for time step $l$ as
\begin{equation}\label{DDM_update_weight}
	{\mathbf{V}}_n^{(l)}\leftarrow\gamma\mathbf{V}_n^{(l\!-\!1)}+(1-\gamma)\mathbf{V}_n^{(l)},
\end{equation}
where $\gamma$ is a hyper-parameter not only used for updating RB decisions but also for preventing gradient vanishing, which typically increases with $L$. At the $L$-th time step, the obtained ${\mathbf{V}}_n^{(L)}$ is the final decision of AP$_n$. 

The sub-DNNs are trained in a centralized manner, while inference is conducted in a distributed manner. The learnable parameters of all sub-DNNs can be updated using the stochastic gradient descent (SGD) algorithm. We employ the sample-averaged negative sum rate as the loss function, which is $\mathcal{L}=-\frac{1}{N_\mathrm{tr}}\sum_{m=1}^{N_\mathrm{tr}}\sum_{f=1}^{F}\sum_{k=1}^{K}\log _2\left( 1+\mathrm{SINR}_{k,f}^{(m)}\right)$, where $N_\mathrm{tr}$ denotes the number of channel samples in a mini-batch. The channels and decisions of all APs (i.e., $\mathbf{V}_1^{(L)}, \cdots, \mathbf{V}_N^{(L)}$) are used to compute the loss function for centralized training. After training, the sub-DNNs are deployed at each AP for distributed inference. The sub-DNN can be achieved through any DNN architectures, such as fully-connected neural networks, convolutional neural networks, or graph neural networks (GNNs).

\section{Simulation Results}

In this section, we evaluate the performance of the proposed algorithms. Although the distributed learning method in \cite{chen2024distributed}, referred to as ``Learn-local'', is not specifically designed to optimize the RB allocation of wideband cell-free system, we can borrow its idea to address the considered problem for comparative purposes. In Learn-local, each AP$_n$ employs local channel $\mathbf{H}_n$ to infer RB decisions using sub-DNN.

\subsection{Simulation Setup}

Consider a $300~\mathrm{m}\times 300~\mathrm{m}$ square area, where 16 APs and 8 UEs (unless otherwise specified) are randomly distributed. For C-SWMMSE, the 16 APs are divided into $Q=4$ clusters, each containing 4 APs. The pathloss follows 3GPP NLoS Urban Microcell model \cite{3gpp36814}, which is $\beta_{n,k}=36.7\log_{10}d_{n,k}+22.7+26\log_{10}f^{\mathsf{c}}$, where $d_{n,k}$ denotes the distance between AP$_n$ (with the height of $10~\mathrm{m}$) and UE$_k$, and $f^{\mathsf{c}}=2~\mathrm{GHz}$ is the carrier frequency. The noise power is given by $\sigma_{k,f}^2=-174+10\log_{10}B+\xi-10\log_{10}F$ dBm, where $F=11$ denotes the number of RBs, $B=10$ MHz is the system bandwidth with a subcarrier spacing of 60 kHz~\cite{3gpp381011}, and $\xi=7$ dB is the noise figure. The small-scale fading channel follows a Rayleigh fading model. We show the sum rate on one subcarrier of each RB (i.e., $\frac{\mathrm{SR}}{C}$) and 2000 channel samples are used for testing. The maximum transmit power of APs and UEs is set to $P_\mathrm{t}=25$ dBm and $P_\mathrm{UE}=23$ dBm, respectively.

We employ GNN as the architecture of sub-DNN, which can exploit the permutation equivariance property of the RB allocation policy to reduce training complexity~\cite{liu2024multi}. Each sub-DNN consists of two 32-neuron hidden layers, while that used in the Learn-local method has four 128-neuron hidden layers. The activation function for the hidden layers is Leaky ReLU. The SGD algorithm is implemented using the Adam optimizer with a learning rate of $1\times 10^{-3}$.

\subsection{Performance Comparison}\label{perf_algo}

\begin{figure}[htbp]
	\centerline{\includegraphics[width=0.8\linewidth]{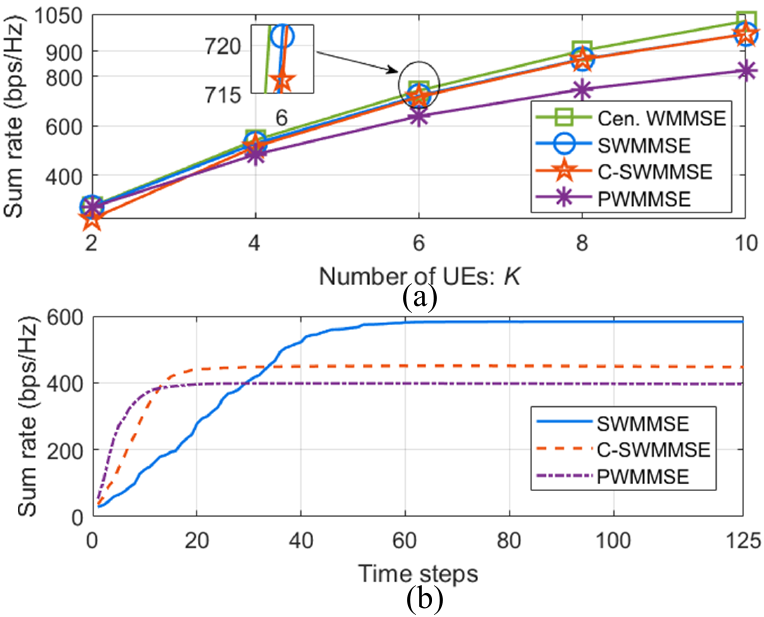}}\vspace{0mm}
	\caption{Sum rate achieved by numerical algorithms. (a) Sum rate at convergence. (b) Convergence curve of SWMMSE, C-SWMMSE, and PWMMSE.}
	\label{cen_seq_wmmse}
\end{figure}

The sum rates achieved by the numerical algorithms are compared in Fig.~\ref{cen_seq_wmmse}(a). The legend ``Cen. WMMSE" stands for the centralized WMMSE. Both SWMMSE and C-SWMMSE achieve sum rates close to that of the centralized WMMSE and higher than that of PWMMSE. The SWMMSE slightly outperforms C-SWMMSE because the outdated information from the APs within the same cluster impairs the performance. The convergence curves are shown in Fig.~\ref{cen_seq_wmmse}(b). While SWMMSE achieves the highest sum rate at convergence, PWMMSE exhibits the fastest convergence speed.

\begin{figure}[htbp]
	\centerline{\includegraphics[width=0.75\linewidth]{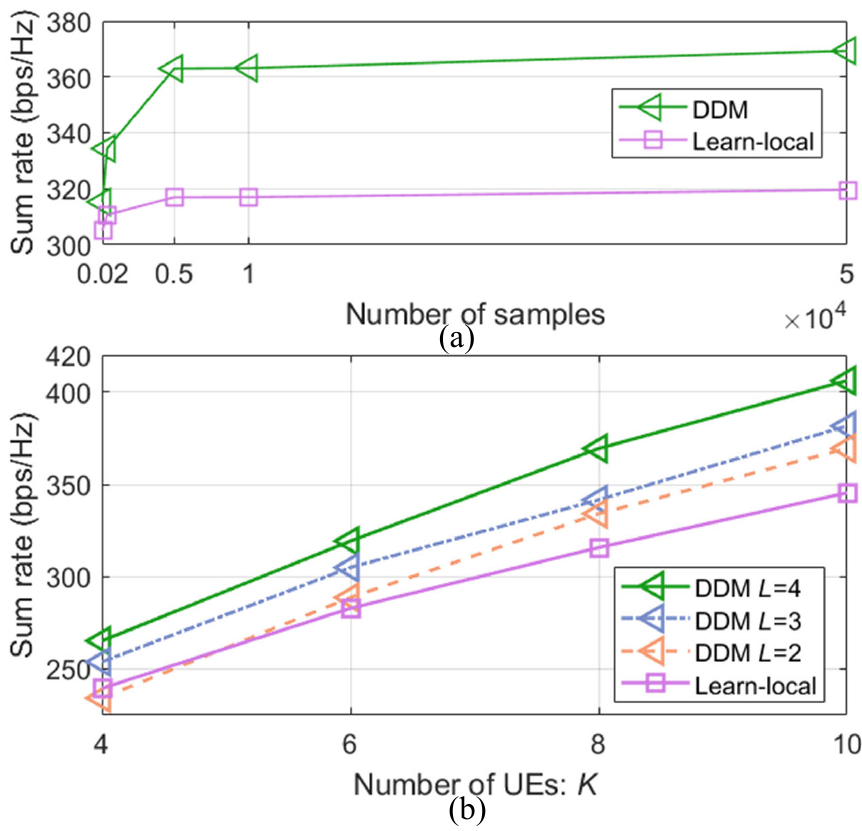}}\vspace{0mm}
	\caption{Comparison between DDM and Learn-local.}
	\label{DDM_CTDE}
\end{figure}

The impact of the number of training samples on DDM and Learn-local is shown in Fig.~\ref{DDM_CTDE}(a). In DDM, APs are given $L=4$ time steps to infer their final decisions. As the number of training samples increases, the sum rate of DDM rapidly improves, whereas the performance gain of Learn-local is modest. In Fig.~\ref{DDM_CTDE}(b), we compare the performance of DDM and Learn-local versus the number of UEs. The results indicate that DDM outperforms Learn-local when given only $L=2$ time steps and the performance gap is enlarged as the number of UEs increases, despite that Learn-local employs sub-DNNs with significantly more learnable parameters than DDM.

\begin{figure}[htbp]
	\centerline{\includegraphics[width=0.8\linewidth]{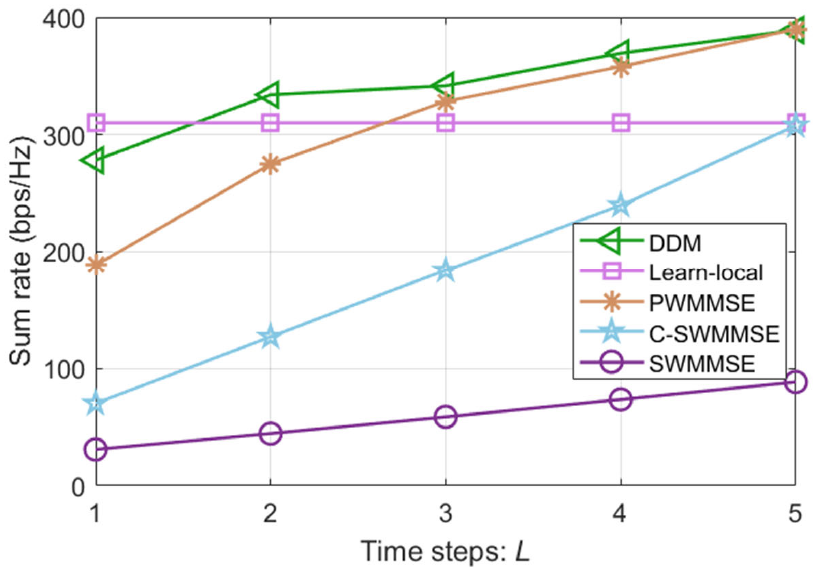}}\vspace{0mm}
	\caption{Performance versus the number of time steps,  50,000 training samples are used for DDM and Learn-local.}
	\label{impact_ts}
\end{figure}

The impact of the number of time steps on different methods is presented in Fig.~\ref{impact_ts}. PWMMSE achieves higher sum rate than SWMMSE and C-SWMMSE. DDM outperforms PWMMSE when $L$ is extremely small. This suggests that DDM learns superior RB allocation compared to that computed by \eqref{how2feedback} when given  extremely limited time steps. As the number of time steps increases, the performance gap between DDM and PWMMSE reduces. The performance of Learn-local remains unchanged because it only utilizes local channel to learn allocation decisions.

\section{Conclusions}

In this paper, we proposed distributed algorithms for optimizing RB allocation of wideband cell-free systems. The proposed algorithms update the decision of each AP locally with the information gathered via OTA transmission. To reduce the signaling overhead associated with OTA transmission, a DL-based method was developed. Simulation results demonstrated that the proposed algorithms achieve a sum rate close to that of the centralized algorithm, and the proposed DDM outperforms an existing distributed learning method and the numerical algorithms when given only a few time steps.

\bibliographystyle{IEEEtran}
\bibliography{Reference}

\end{document}